# AN ALTERNATE DESIGN FOR CLIC MAIN LINAC WAKEFIELD SUPPRESSION

V. F. Khan and R.M. Jones; Cockcroft Institute, Daresbury, WA4 4AD, UK;
University of Manchester, Manchester, M13 9PL, UK.

*Abstract*

The present design of the main accelerating structure for CLIC is based on heavy damping (WDS) with a Q of ~10. The wakefield suppression in this case entails locating the damping materials in relatively close proximity to the accelerating cells. Herein we present an alternate design for the main accelerating structures. We detune the lowest dipole band by prescribing a Gaussian distribution to the cell parameters and consider moderate damping Q~500 to prevent the recoherence of the modes; in this case the damping materials can be located at an extended distance from the accelerating structure. The procedure to achieve a well-damped wakefield is described. Results are presented elucidating the various designs including the current one which is being developed to incorporate r.f. breakdown, pulse surface heating and beam dynamics constraints.

## INTRODUCTION

The CLIC scheme is being developing to achieve a 100 MV/m average accelerating gradient at an operating frequency of 12 GHz in order to collide electrons and positrons at 3 TeV centre of mass energy. The present baseline design for CLIC relies on heavy damping (with Qs as low as 10) in order to suppress wake-fields and is referred to as the CLIC_G structure [1], which in essence is a damped waveguide design. CLIC_G has an overall r.f. to beam efficiency of 27.7 % [1] and is designed with a linear tapering of the cell parameters to achieve the required high gradient. This tapering also spreads the dipole frequencies which helps in suppressing the wake-fields; however for a given bandwidth (~1GHz) wakefield suppression in this design is insufficient and heavy damping is required Q~10.

In order to achieve this heavy damping the dimensions of the damping waveguides need to be comparatively large and the damping material may end up being located relatively close to the accelerating cells. This will have an impact on the fundamental mode properties.

In order to isolate the fundamental mode properties of the accelerating cells from the damping materials we consider an alternative scheme which relies on the detuning of the lowest dipole modes and moderate damping with Q~500. In a moderate Q scheme the damping materials can be located remotely from the accelerating cells. Just such a scheme was developed for the NLC design, which was operated at the X-band accelerating frequency, but with an accelerating gradient of ~ 65 MV/m [2].

For a moderate damping scheme for the CLIC main accelerating linac we considered several design structures. It has been observed that larger bandwidth (of lowest dipole mode) structures satisfy the beam dynamics constraints, but do not satisfies the r.f. breakdown constraints. For smaller bandwidth structures both these constraints are satisfied, but the r.f. to beam efficiency is very poor. In order to obtain better r.f. to beam efficiency we consider the possibility of a zero crossing structure. In this paper we present the results of a zero crossing structure which satisfies both constraints with an r.f. to beam efficiency of 26.1%.

## CONSTRAINTS

1. R.F. breakdown constraints [1]: R.F. constraints are given by r.f. breakdown and pulse surface heating constraints:

a. Surface electric field: $E_{sur}^{max}$ < 260 MV/m
b. Pulsed surface heating: $\Delta T^{max}$ < 56 K
c. Power: $P_{in}\tau_p^{1/3}/C_{in}$ < 18 MW ns$^{1/3}$/mm

Here $E_{sur}^{max}$ and $\Delta T^{max}$ refer to maximum surface electric field and maximum pulse surface heating temperature rise in the structure, respectively. If the surface electric field exceeds the above limit there is the possibility that electrons will be pulled from the cavity surface (Cu) leading to breakdown. Pulse surface heating is proportional to the square of the surface magnetic field [3], if the pulse temperature exceeds the above limit surface damage due to thermally induced stresses or (in the extreme case) melting of the cavity walls may occur. $P_{in}$ and $\tau_p$ denotes the input power and pulse length respectively and $C_{in}$ is the iris circumference of the first cell. In order to optimise the structure the cost factor constraint (i.e. 1c) should also be satisfied.

2. Beam dynamics constraints [4],[5]: Beam dynamics constraints are determined by considering emittance growth due to short and long range transverse wakefields:

a) N – bunch population is decided by the structure parameters $<a>/\lambda$, $\Delta a/<a>$, this is because the short range wake is inversely proportional to the fourth power of the average iris radius [6]. Here $<a>$ is an average of the iris radii of the end cells, $\Delta a$ is half the difference of the iris radii of the end cells and $\lambda$ is the accelerating wavelength.

b) Bunch separation depends on the long range transverse wakes and is determined by the condition:

$$W_{t1} < \frac{10 \text{ V/pC/mm/m} \times 4 \times 10^9 \times <E_{acc}>}{N \times 150 \text{ MV/m}} \quad [4]$$

$W_{t1}$ is the transverse wake on the first trailing bunch and $<E_{acc}>$ is the average accelerating gradient. The beam dynamics constraints indicate that in order to reduce the short range wakefield effect as the average iris radius is reduced the number of particles per bunch must be limited. For a given structure with $<.E_{acc}> = 100$ MV/m if $N \sim 4 \times 10^9$ then bunch separation should be such that transverse wake on the first trailing bunch should be less than 6.7 V/pC/mm/m and the rest bunches should see a wake less than that of first trailing bunch.

## STRUCTURE COMPARISION

The nominal bunch spacing for the present CLIC_G structure is 6 cycles (0.5 ns) [1]. In our previous work we focussed only on wakefield issues and thus we explored the bandwidth regime so as to suppress the wakefield for a 0.5 ns bunch spacing below the acceptably set limit, in which it was observed that for a bandwidth of 3.3 GHz the wakefield is adequately suppressed. The methodology and results of this Gaussian detuning of the cell parameters, both with and without interleaving structures is explained in detail in [7]. Unfortunately the surface fields of this design are very high and it would be difficult to satisfy r.f. constraints for this particular structure.

In [8] we explained the methodology of a possible zero crossing scheme with parameters closely matched to that of CLIC_G structure. The purpose of looking into a zero crossing scheme is to satisfy the rf breakdown as well as beam dynamics constraints without affecting the r.f. to beam efficiency of the collider. In the CLIC_ZC structure [9] (now referred to as CLIC_ZC1), the surface electric field is ~10.0 % above the limit and hence fails to satisfy the r.f. breakdown criteria. We are able to reduce the surface field by changing the ellipticity of the cells, for the current structure (CLIC_ZC2) the ratio of the major to the minor ellipse axis is 1:1.3. The parameters of 5 fiducial cells of a 24 cell structure which are referred to as the CLIC_ZC2 structure are presented in Table 1, where a is iris radius, b is cavity radius, t is iris thickness, Vg/c is group velocity of the fundamental mode and $\omega_1/2\pi$ is the lowest dipole synchronous frequency.

The fundamental mode properties of the 3 designs are compared in Table 2. The advantage of a small bandwidth (~1 GHz) structure is the shunt impedance (R')of the cells are comparatively high, this is because as the iris radius of the cells is reduced the corresponding transit time factor improves; such structures requires less power to achieve the desired average accelerating gradient and constraint 1 is satisfied. The disadvantage is that the Q of the dipole modes needs to be low enough to suppress the wakefields, to satisfy beam dynamics criteria. The variation of the surface fields, power and the accelerating field as a function of distance is illustrated in Fig. 1.

Table 1: The parameters of 5 of the fiducial cells of the CLIC_ZC2 structure.

| Cell No. | a (mm) | b (mm) | t (mm) | Vg/c (%) | $\omega_1/2\pi$ (GHz) |
|---|---|---|---|---|---|
| 1 | 2.87 | 9.89 | 1.6 | 1.45 | 17.56 |
| 6 | 2.69 | 9.83 | 1.38 | 1.25 | 17.79 |
| 12 | 2.55 | 9.78 | 1.2 | 1.12 | 17.96 |
| 18 | 2.40 | 9.74 | 1.04 | 1.0 | 18.13 |
| 24 | 2.13 | 9.68 | 0.7 | 0.83 | 18.40 |

Table 2: The parameters of the several CLIC structures (IP = Input i.e. first cells, OP = Output i.e. last cell)

| Structure | CLIC_G [1] | CLIC_ZC1[9] | CLIC_ZC2 |
|---|---|---|---|
| $<a>/\lambda$ | 0.11 | 0.102 | 0.1 |
| IP /OP  a (mm) | 3.15, 2.35 | 2.99/2.13 | 2.87/2.13 |
| IP /OP  t (mm) | 1.67, 1.0 | 1.6/0.7 | 1.6/0.7 |
| IP /OP vg/c (%) | 1.66/0.83 | 1.49/0.83 | 1.45/0.83 |
| IP /OP  Q | 6100/6265 | 6366/6643 | 6408/6668 |
| IP /OP R' (MΩ/m) | 89/112 | 107 / 138 | 108/138 |
| Filling time (ns) | 62.9 | 56.8 | 58.6 |
| $P_{in}$ (MW) | 63.8 | 48.0 | 47.0 |
| RF-to-beam efficiency (%) | 27.7 | 27.09 | 26.11 |
| No. of bunches | 312 | 312 | 312 |
| Bunch population | $3.7\times10^9$ | $3.0\times10^9$ | $2.9\times10^9$ |
| $E_{sur}^{max}$ (MV/m) | 245 | 285 | 231 |
| $\Delta T^{max}$ (K) | 53 | 20.0 | 25.2 |
| $P_{in}\tau_p^{1/3}/C_{in}$ (MW ns$^{1/3}$/mm) | 18.0 | 14.07 | 14.36 |
| FOM (a.u.) | 9.1 | 8.47 | 8.03 |

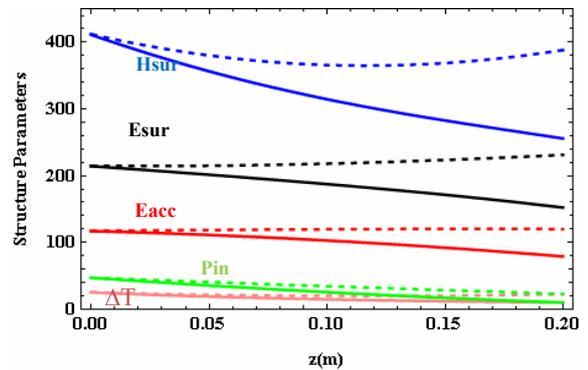

Figure 1: Fundamental mode properties of the CLIC_ZC2 structure. The curves are surface magnetic field (1000*A/m)[blue], surface electric field (MV/m)[black], accelerating gradient (MV/m)[red],  power (MW)[green] and pulse surface temperature rise (K)[pink]. Dashed curves are unloaded and solid curves are beam loaded conditions.

It should be emphasised that the pulse temperature rise in the CLIC_G structure is high because it is heavily damped, with four waveguides connected to each cell.

However, both versions of the CLIC_ZC structures are detuned and moderately damped, hence the pulse temperature rise is comparatively less than that of CLIC_G. In future work we will incorporate four manifolds [9] connected to the cells to remove the higher order modes; due to the inclusion of the manifold geometry we expect pulse temperature rise to increase by a factor of ~2 [10].

## WAKEFIELD IN CLIC_ZC2 STRUCTURE

As the dimensions of both the ZC structures are closely matched, the envelope and the amplitude of the wakefiled for these structures are very similar. The resulting wake-field amplitude plot for the 24 cell CLIC_ZC2 structure is illustrated in Fig.2 with a damping Q = 500.

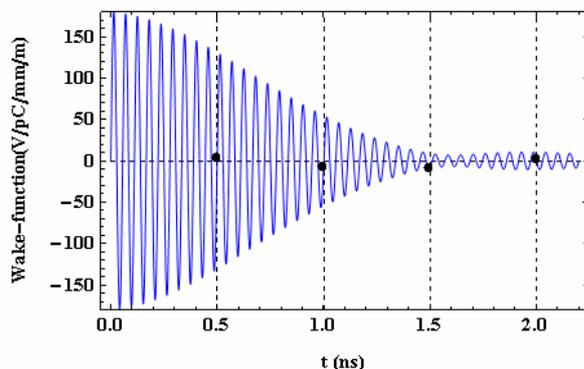

Figure 2: Amplitude of wakefield in the CLIC_ZC2 structure representing the wakefield at first four trailing bunches. The wakefield at the location of the bunch is represented by the dots.

The wakefield suppression is further improved when mode frequencies of successive structures are interleaved and this is illustrated in Fig.3.

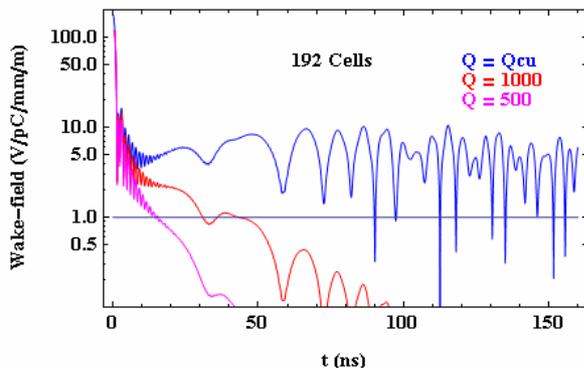

Figure 3: Envelope of wakefield for an 8 fold interleaving of dipole mode frequencies in the CLIC_ZC2 structure.

The sensitivity of the RMS of the sum wake-field ($S_{RMS}$) to small fractional errors in the bunch spacing or in the mode frequencies provides an indication where BBU occurs [11]. The $S_{RMS}$ for a 24 cell structure is illustrated in Fig 3 (inset), as it is seen in Fig. 3, for a nominal bunch spacing $S_{RMS}$ is ~30 V/pC/mm/m which is above the acceptable limit of unity [11], but for an interleaved structure it is reduced to ~7 V/pC/mm/m which unfortunately is still above the unity.

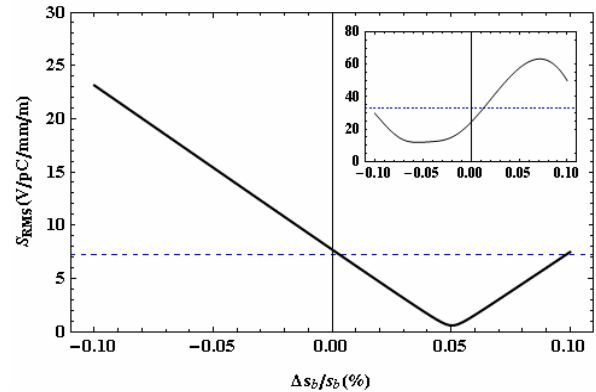

Figure 4: $S_{RMS}$ of a 192 cell structure for the CLIC_ZC2 design with a damping Q = 500, the 24 cell structure result shown in the inset.

## DISCUSSION

After considering several structures it is observed that smaller bandwidth structures will satisfy r.f. as well as beam dynamics constraints; the additional condition of having a zero crossing in such designs will maintain the efficiency of the collider. However, the zero crossing condition is very sensitive to fabrication errors. Further study is ongoing to optimise the zero crossing condition in which fabrication tolerances and the damping manifold geometry has been taken into account.

## ACKNOWLEDGMENTS

We have benefited from valuable discussions with W. Wuensch and A. Grudiev regarding structure optimisation and with D. Schulte on the beam dynamics issues. We are pleased to acknowledge many useful discussions with I. Shinton, C. Glasman and N. Juntong.